\documentclass[a4paper]{jpconf}
\usepackage{graphicx}
\usepackage{xcolor}
\usepackage{iopams}
\usepackage{array,multirow}
\usepackage{listings}
\usepackage{comment}

\lstdefinestyle{mlstyle}{
    escapechar={!},
    language=C++,
    keywordstyle=\color{purple},
    commentstyle=\color{commentgreen},
    stringstyle=\color{magenta},
    numberstyle=\tiny\color{gray},
    basicstyle={\footnotesize\ttfamily},
    frame=single,
    breaklines=true,
    captionpos=b,
    keepspaces=true,
    numbersep=5pt,
    showspaces=false,
    showstringspaces=false,
    showtabs=false,
    tabsize=2,
    firstnumber=1,
    numberfirstline=true,
    otherkeywords={OFFLOAD,\_\_global\_\_,pragma},
    morekeywords=[2]{OFFLOAD,\_\_global\_\_},
    morekeywords=[3]{loop, independent, target, teams, device, pragma, acc, omp, parallel, copyin, copyout, mapto, mapfrom, map, distribute, for},
}

\begin{document}
\title{Portable Programming Model Exploration for LArTPC Simulation in a Heterogeneous Computing Environment: OpenMP vs. SYCL}

\author{
Meifeng Lin$^1$, 
Zhihua Dong$^1$, 
Tianle Wang$^1$, 
Mohammad Atif$^1$, 
Meghna Battacharya$^2$,
Kyle Knoepfel$^2$,
Charles Leggett$^3$, 
Brett Viren$^1$, 
Haiwang Yu$^1$
}

\address{$^1$ Brookhaven National Laboratory, Upton, NY 11973, USA }
\address{$^2$ Fermi National Accelerator Laboratory, Batavia, IL 60510, USA }
\address{$^3$ Lawrence Berkeley National Laboratory, Berkeley, CA 94720, USA }



\newcommand{\meifeng}[1]{\textcolor{red}{#1}}

\begin{abstract}
The evolution of the computing landscape has resulted in the proliferation of diverse hardware architectures, with different flavors of GPUs and other compute accelerators becoming more widely available. To facilitate the efficient use of these architectures in a heterogeneous computing environment, several programming models are available to enable portability and performance across different computing systems, such as Kokkos, SYCL, OpenMP and others. As part of the High Energy Physics Center for Computational Excellence (HEP-CCE) project, we investigate if and how these different programming models may be suitable for experimental HEP workflows through a few representative use cases. One of such use cases is the Liquid Argon Time Projection Chamber (LArTPC) simulation which is essential for LArTPC detector design, validation and data analysis. Following up on our previous investigations of using Kokkos to port LArTPC simulation in the Wire-Cell Toolkit (WCT) to GPUs, we have explored OpenMP and SYCL as potential portable programming models for WCT, with the goal to make diverse computing resources accessible to the LArTPC simulations. In this work, we describe how we utilize relevant features of OpenMP and SYCL for the LArTPC simulation module in WCT. We also show performance benchmark results on multi-core CPUs, NVIDIA and AMD GPUs for both the OpenMP and the SYCL implementations. Comparisons with different compilers will also be given where appropriate. 
\end{abstract}


\section{Introduction} 

Computing plays a central role in many aspects of experimental high energy physics (HEP), including detector simulation, event reconstruction, and end-user data analysis. Traditionally, experimental HEP software is written with homogeneous multi-core CPUs as the main targets. 
 With the expected orders-of-magnitude increase of data volume and data rate from the next-generation high-energy particle experiments, purely relying on CPU computing resources may not be sufficient to meet the demands of the processing needs. One such experiment is the planned Long-Baseline Neutrino Facility (LBNF)/Deep Underground Neutrino Experiment (DUNE), an international project to study the fundamental properties of neutrinos. Built on the Liquid Argon Time Projection Chamber (LArTPC) technology, DUNE is expected to bring 30 PB per year of data from the far detector (FD) LArTPC modules alone~\cite{DUNE:2022fcw}. This is at least an order of magnitude more than that from the current and past neutrino experiments. A single readout of one of the four FD modules is expected to provide about 5 GB of raw packed data at a rate of about 0.1 Hz. In addition, roughly 500 TB of data is expected to arrive from the FD each month on average due to observing activity consistent with that of a supernova neutrino burst.  Though the vast majority of these triggers will be due to background fluctuations, each one must nonetheless be fully analyzed. Analyzing DUNE FD data promptly may require access to and efficient utilization of modern high performance computing (HPC) systems, which often have heterogeneous configurations with host multi-core CPUs plus compute accelerators such as general-purpose Graphics Processing Units (GPUs). To efficiently utilize these HPC resources, existing software will need to be refactored to allow certain computation-intensive parts of the workload to run on GPUs. 
 However, as HEP software needs to go through rigorous verification and validation processes, we do not want to and should not rewrite the software for every new hardware architecture that comes out. This is especially important now as there are at least three different GPU vendors, AMD, Intel, and NVIDIA, who have different native application programming interfaces (APIs) for their GPUs. 

We are thus motivated to explore portable programming models that would allow us to write the software against a single API while delivering decent performance over a variety of CPU and GPU architectures. This is part of the strategy pursued by the High Energy Physics Center for Computational Excellence (HEP-CCE) project, which investigates ways to get HEP workflows to run efficiently on various HPC systems. A test use case that is particularly of interest to DUNE is the LArTPC signal simulation, which models the electric current induced in the LArTPC wires by ionization electrons that are produced by energetic charged particles passing through the detector. In previous studies~\cite{yu2021evaluation,dong2023evaluation}, we have shown that Kokkos~\cite{edwards2014kokkos}, a C++ abstraction layer that supports multiple architectures, can provide performance and portability while maintaining the familiar C++ syntax and ease of use. Here we present our experiences in porting the LArTPC signal simulation to two other portable programming models, SYCL~\cite{reyes2016sycl} and OpenMP~\cite{openmp}. We briefly describe LArTPC signal simulation in the Wire-Cell Toolkit (WCT) in Section~\ref{sec:wct}. A high-level description of Kokkos, SYCL, and OpenMP is given in Section~\ref{sec:comp}. The SYCL and OpenMP implementation details are presented in Section~\ref{sec:impl}. In Section~\ref{sec:eval}, we document the performance benchmarks for Kokkos, SYCL, and OpenMP. And we summarize our experiences in Section~\ref{sec:summary}. 

\section{LArTPC Signal Simulation in the Wire-Cell Toolkit}  \label{sec:wct}


The Wire-Cell Toolkit (WCT)~\cite{wct} currently provides state-of-the-art LArTPC signal simulation implementation. The LArTPC signal simulation involves three main steps, as described in more detail in Ref.~\cite{yu2021evaluation}:
\begin{enumerate}
\item \textbf{Rasterization}: Individual ionization electron groups are binned into ``patches'' of varying sizes, typically with dimensions of $\mathcal{O}(20\times20)$ elements. This is where most of the floating point operations occur and can benefit greatly from parallelization, since operations over different patches can be done in parallel. 

\item \textbf{Scatter-Add}: These patches are stacked and summed over a larger grid that spans the longitudinal time and transverse space of the union of the patches. The grid is typically an array of $\mathcal{O}(10,000 \times 10,000)$ elements. This step requires the \texttt{atomic add} operation.

\item \textbf{Convolution}: The large grid is then convolved with the detector response with dimensions $\mathcal{O}(100 \times 100)$ to obtain the simulated signal. The main computation here is Fast Fourier Transform (FFT).
\end{enumerate}
To simplify the testing of different portable programming models, we developed a standalone code, \texttt{wire-cell-gen}, for the LArTPC signal simulation based on WCT, and implemented Kokkos, SYCL and OpenMP versions separately~\cite{wct-kokkos, wct-sycl, wct-omp}. For each implementation, the above three steps are all contained in the \texttt{DepoTransform} function. 

\section{Kokkos, SYCL, and OpenMP} \label{sec:comp} 
Kokkos~\cite{edwards2014kokkos} is a C++ abstraction layer that provides architecture-agnostic APIs for portability. It has gained popularity since its inception due to its support for multiple hardware architectures, including multicore/many-core CPUs,  NVIDIA GPUS, AMD GPUs, and Intel GPUs. 

SYCL~\cite{sycl} is an open standard programming model developed by the Khronos Group, which enables writing  C++ applications that can run on heterogeneous systems, such as CPUs, GPUs,  and FPGAs etc.  Developers can write single source code and build with supported compilers for different backends targeting  different hardware platforms. It is also possible to select target at runtime if it is supported by the compiler. 

OpenMP~\cite{openmp} is an API that include compiler directives and runtime library routines for multithreading on CPUs and “target offloading” on heterogeneous architectures. 
With the appropriate compilers, an OpenMP code can potentially run on CPUs, GPUs and FPGAs. 

\section{Implementation Details} \label{sec:impl}
\subsection{Wrappers for Vendor-Optimized FFT and RNGs}
Since \texttt{wire-cell-gen} uses FFT and random number generators (RNGs), which can be computationally expensive, we want to be able to use the optimized libraries provided by the hardware vendors. Currently, Kokkos, SYCL, and OpenMP do not provide a mechanism to use these optimized libraries in a portable way. As such, similar to our Kokkos implementation~\cite{dong2023evaluation}, we write our own wrappers to allow us to use vendor-optimized libraries for FFT and RNGs.  For FFT, we use \texttt{cuFFT} for NVIDIA GPUs and \texttt{rocFFT} for AMD GPUs. For the host CPU code, we use non-parallel \texttt{fftw} as it is used in the original WCT code.  For RNGs, we use \texttt{cuRAND} for NVIDIA GPUs, \texttt{rocRAND} for AMD GPUs  and use \texttt{random123}~\cite{random123,salmon2011parallel} for CPUs. 

\subsection{SYCL}
In Ref.~\cite{dong2023evaluation} we described our procedure and performance of porting the \texttt{wire-cell-gen} code to Kokkos. Since we had already structured the data flow and parallel execution patterns to be GPU-friendly in the Kokkos implementation, we started the SYCL implementation from the Kokkos version. In many cases, the SYCL syntax is very similar to Kokkos, so porting from Kokkos to SYCL is relatively straightforward.



We created \texttt{Array1D} and \texttt{Array2D} classes to replace \texttt{KokkosArray} class in the Kokkos implementation (which is a wrapper for \texttt{KokkosView}).  
Those two classes have just simple pointers and sizes with a few methods. Using them,  many lines of Kokkos code can be easily converted to SYCL with simple translations such as shown below.

\begin{verbatim}
Kokkos::deep_copy(sps_f, spf_h);   ->  sp_fs.copy_from(sps_h);
auto  sp_ts =                      ->  auto  sp_ts = 
KokkosArray::idft_cr(sp_fs,1) ;        SyclArray::idft_cr(sp_fs,1) ;
\end{verbatim}



\subsection{OpenMP} 
For many CPU-based projects, porting using OpenMP means we can start by adding \verb|#pragma omp| for data movement and loop parallelization, and do not need to change the CPU code significantly. However, the original WCT code is not structured well for this form of parallelization and so we started from the Kokkos version. We used one-dimensional arrays to represent all the data. We also manually performed data movement using \verb|#pragma omp| target data map to remove unnecessary data movement and lower peak memory usage. 

One key difference between the Kokkos version and the OpenMP version is that, since Kokkos supports GPU prefix sum operation (scan) while OpenMP currently does not, we modified the algorithm for one of the kernels to remove the use of GPU scan operation. This change only introduces negligible performance loss and relatively small extra memory usage.



\section{Performance Evaluations} \label{sec:eval}

\subsection{Hardware Platform and Software Environment}
For consistency of performance benchmarks, we used the same workstation (lambda1) at Brookhaven National Laboratory which has an NVIDIA V100 GPU, an AMD Raedon Pro VII GPU, and an AMD 24-core Ryzen Threadripper 3960X CPU with 48 hyperthreads. This is the same platform used for the performance evaluation of the Kokkos implementation~\cite{dong2023evaluation}.


The compiler support for both SYCL and OpenMP target offloading is still under constant development. We  tried a few SYCL compilers, including Intel LLVM compiler, OpenSYCL (formerly hipSYCL) and Intel oneAPI compiler (\texttt{dpcpp}). None of them was able to successfully build our code  for all backends.  But we were able to have at least one compiler that worked for a particular target backend. The situation is better for OpenMP, as the LLVM Clang compiler works well for all backends we tested. We also tried GCC and NVIDIA's HPC SDK (\texttt{nvc++}) for OpenMP target offloading. NVIDIA HPC SDK only works for NVIDIA GPUs. As for GCC, it can compile and run various unit tests successfully with OpenMP target offloading. However we encountered some runtime errors when trying to compile and run  \texttt{wire-cell-gen-openmp} with GCC. The compilers that  can successfully build our code are summarized in Table~\ref{tab:compilers}. 


\begin{table}[htbp]
    \centering
    \begin{tabular}{|c|c|c|c|}
    \hline
    Programming Model & Compilers & Version & Target Architecture    \\ 
    \hline
   \multirow{4}{*}{SYCL} & \multirow{2}{*}{Intel/LLVM (clang++)} & \multirow{2}{*}{nightly-20220425}&  NVIDIA GPU \\ 
     \cline{4-4}
     &  & & AMD GPU \\
      \cline{2-4}
     &  OpenSYCL  & v0.93 &  CPU     \\
      \cline{2-4}
     &  Intel oneAPI (dpcpp) & 2022.02 & CPU     \\
     \hline
      \multirow{4}{*}{OpenMP}  &  \multirow{3}{*}{LLVM/Clang}  & \multirow{3}{*}{15} &  NVIDIA GPU  \\ 
    \cline{4-4}
     & & & AMD GPU  \\
    \cline{4-4}
     & &  & CPU  \\
    \cline{2-4}
      & NVIDIA HPC SDK (nvc++) & 21.9 &  NVIDIA GPU     \\
\hline
    \end{tabular}
    \caption{ SYCL and OpenMP compilers used for this study. }
    \label{tab:compilers}
    \end{table}

\subsection{Performance Comparison} 
\begin{figure}
    \centering
    \begin{minipage}{0.48\textwidth}
        \centering
        \includegraphics[width=1.0\textwidth]{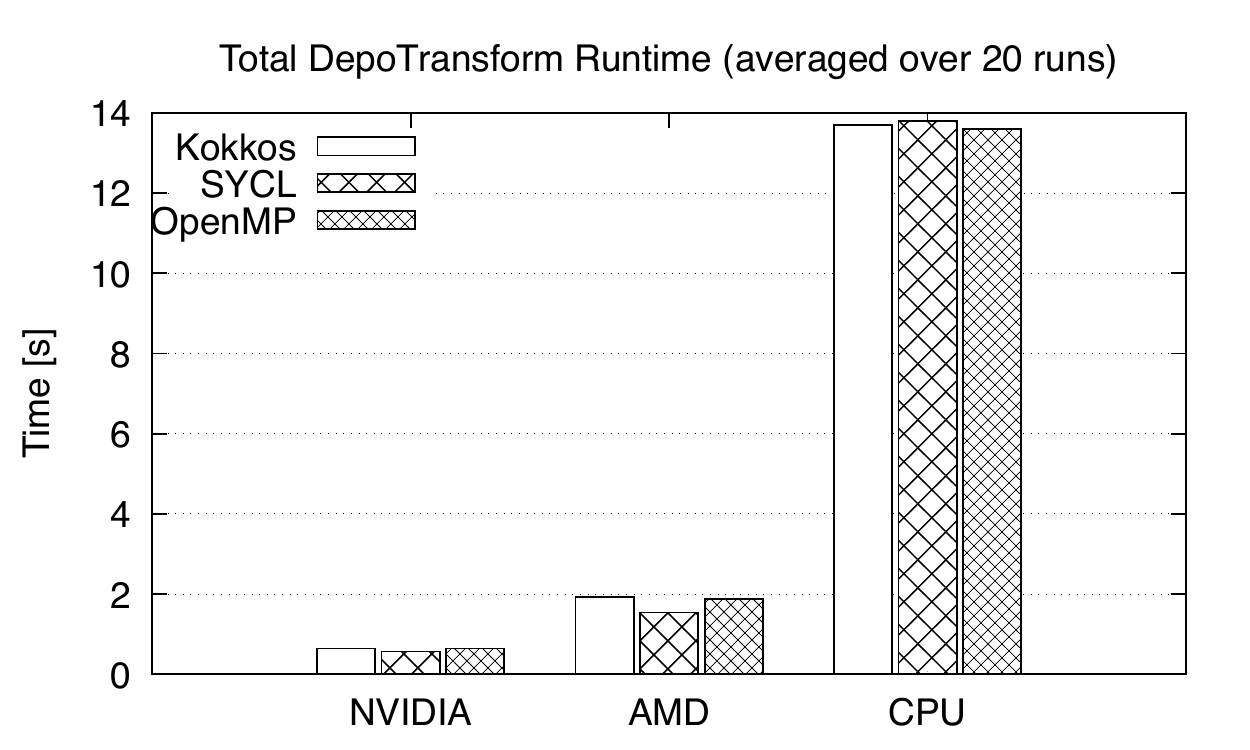} 
         \caption{Comparison of total DepoTransform run time.}
\label{fig:total}
    \end{minipage}\hfill
    \begin{minipage}{0.48\textwidth}
        \centering
        \includegraphics[width=1.0\textwidth]{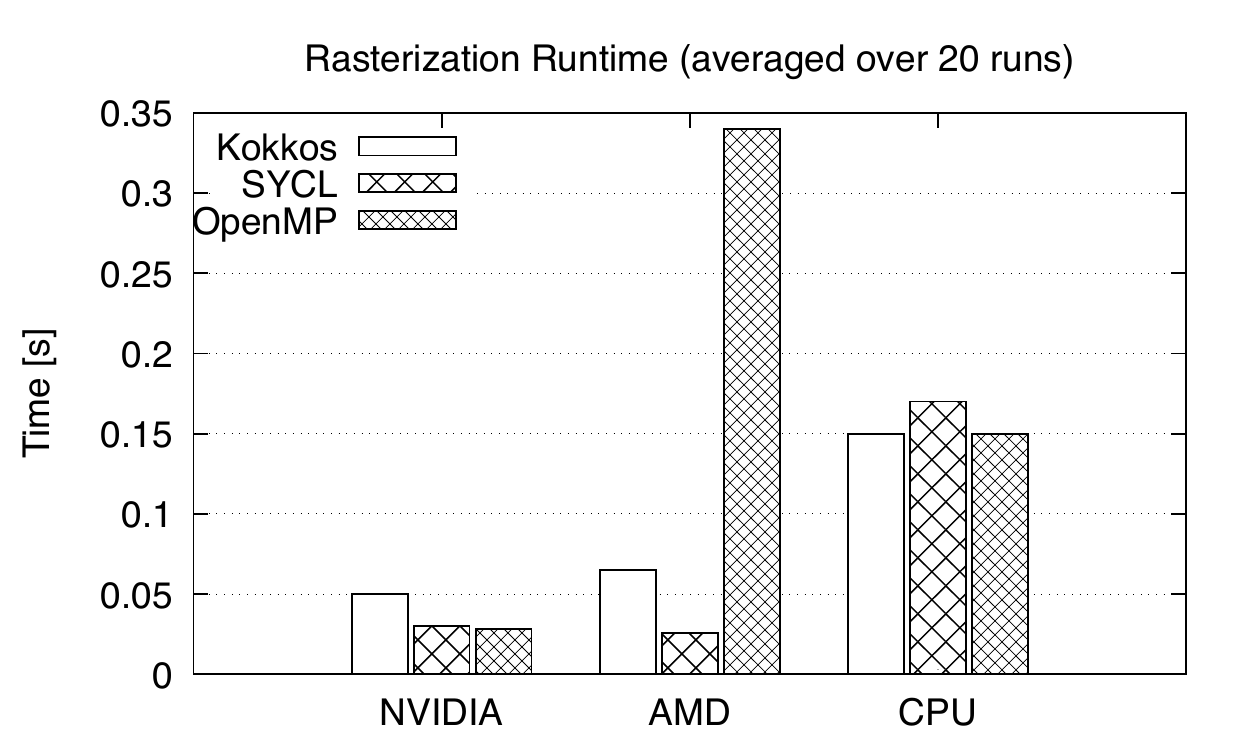} 
        \caption{Comparison of time spent in the Rasterization kernel.}
        \label{fig:rast}
    \end{minipage}
    \begin{minipage}{0.48\textwidth}
        \centering
        \includegraphics[width=1.0\textwidth]{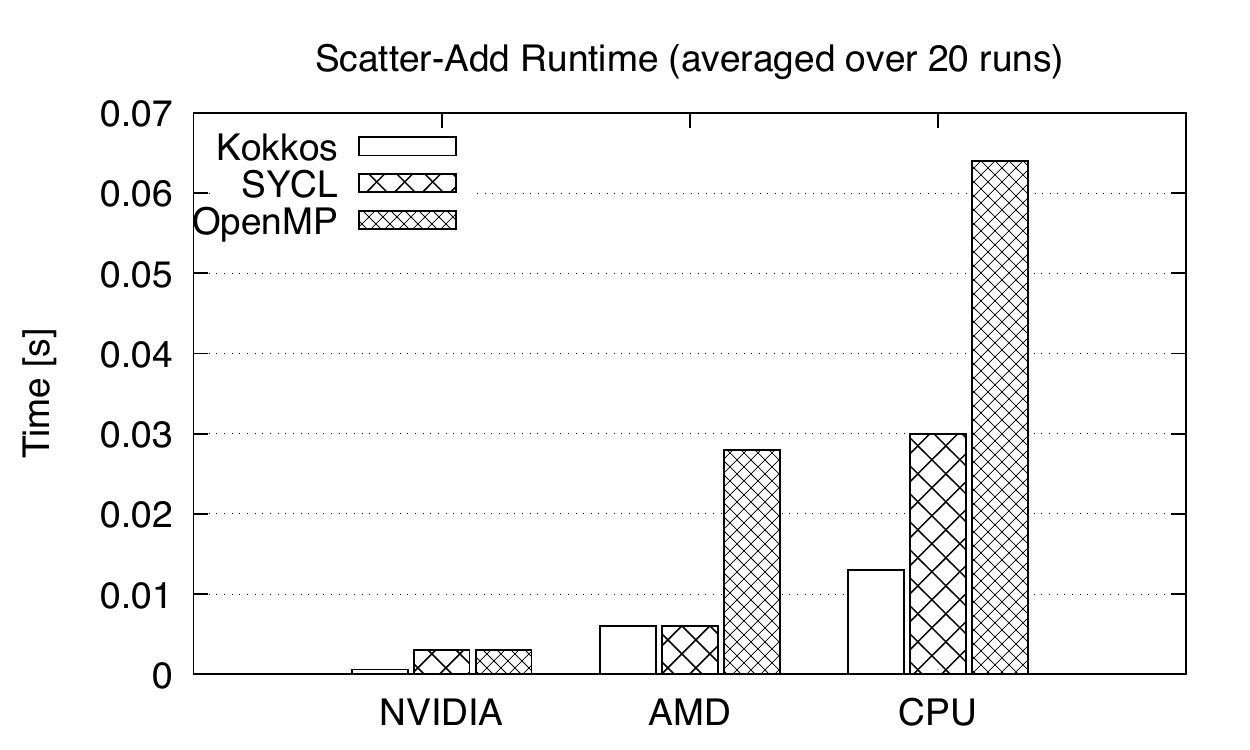} 
         \caption{Comparison of time spent in the Scatter-Add kernel.}
        \label{fig:scadd}
    \end{minipage}\hfill
    \begin{minipage}{0.48\textwidth}
        \centering
        \includegraphics[width=1.0\textwidth]{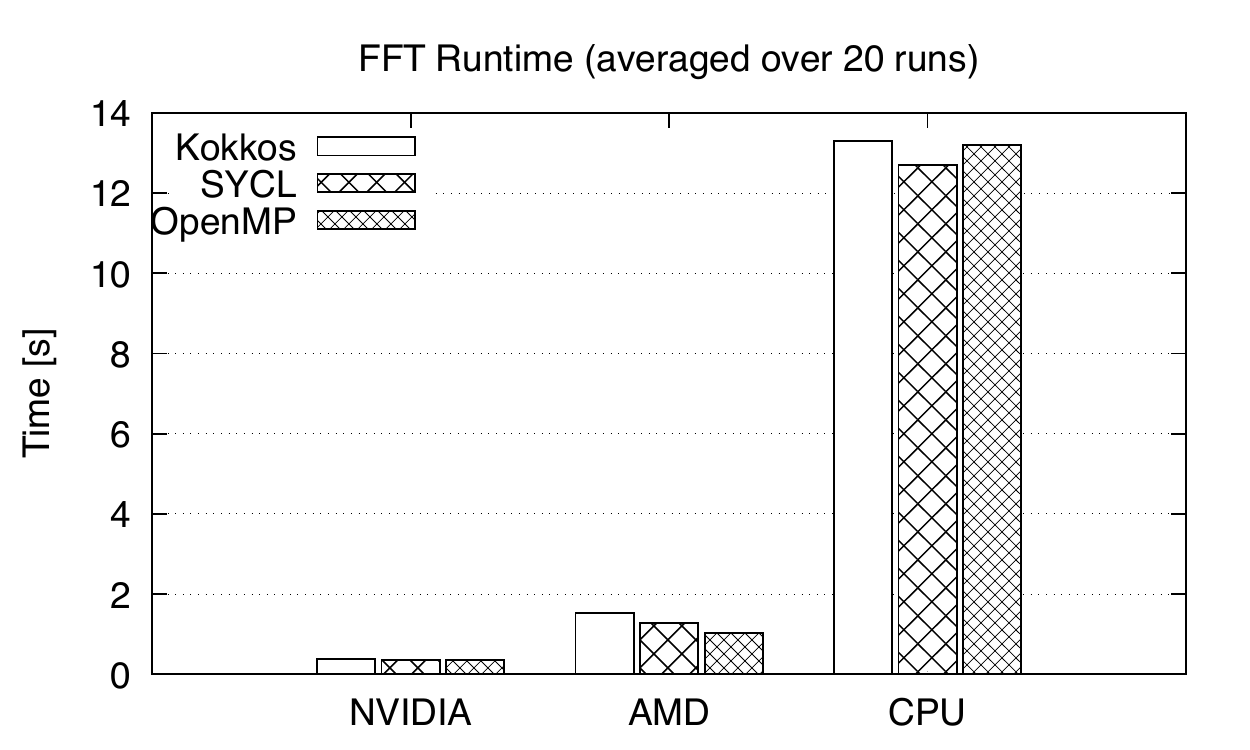} 
        \caption{Comparison of time spent in FFT.}
        \label{fig:fft}
    \end{minipage}
\end{figure}

The SYCL and OpenMP implementations were validated against the Kokkos version for correctness. To compare the performance of each of our implementations on different architectures, we measured the wall-clock time of the key computational kernels, \texttt{Rasterization}, \texttt{Scatter-Add}, and \texttt{FFT}, as well as the total \texttt{DepoTransform} time. The timing measurements were averaged over 20 runs for each  implementation/backend. Figures~\ref{fig:total} to \ref{fig:fft} show the timing results for each backend, where for the SYCL CPU backend, the results are from the oneAPI compiler, and for the OpenMP NVIDIA GPU backend, the results are from the LLVM/Clang compiler. We also include the timing results for the Kokkos implementation~\cite{dong2023evaluation} for comparison. 

Figure~\ref{fig:total} shows that on the same platform, total running times are comparable between different programming models with less than $20\%$ difference. 
However, we can see from the decomposition that the performances of three programming models are very different for the Scatter-Add kernel (Figure~\ref{fig:scadd}), which is based on the \texttt{atomic} operation. The Kokkos implementation gives the best performance, while the OpenMP implementation is about 5-10 times slower than that. In addition, on the AMD GPU, the OpenMP implementation has a large overhead with the first RNG call, resulting in the large timing difference for the Rasterization kernel (Figure~\ref{fig:rast}). For FFT (Figure~\ref{fig:fft}), since different implementations all use the same optimized FFT libraries, the performance difference is small. 

\subsection{Performance Tuning} 
To get the performance results shown above, we went through some performance tuning for both the SYCL and OpenMP implementations. For example, in the OpenMP implementation, the default choice of \verb|num_threads| and \verb|num_teams| sometimes do not offer  the best performance. Another example is that, in Rasterization, there are several kernels with 2-level nested loops of a larger outer loop and a smaller inner loop. We find that for some kernels it is best to parallelize both loops, while in some other cases, it is best to only parallelize the outer loop. For the SYCL CPU backend, there are two ways to compile the program using \texttt{dpcpp}: Just-In-Time (JIT) compilation, or Ahead-Of-Time (AOT) compilation. JIT compilation is the default when the CPU target is \textit{not} explicitly specified through the \texttt{-fsycl -fsycl-targets=spir64\_x86\_64 -Xs} flags. In the case of WCT, JIT adds 1 second to the timing of Rasterization compared to AOT for the CPU backend. We only show the SYCL AOT results in Figures~\ref{fig:total}--\ref{fig:fft}. 

\subsection{Compiler Comparison}
\begin{figure}
    \centering
   \begin{minipage}{0.48\textwidth}
        \centering
        \includegraphics[width=1.0\textwidth]{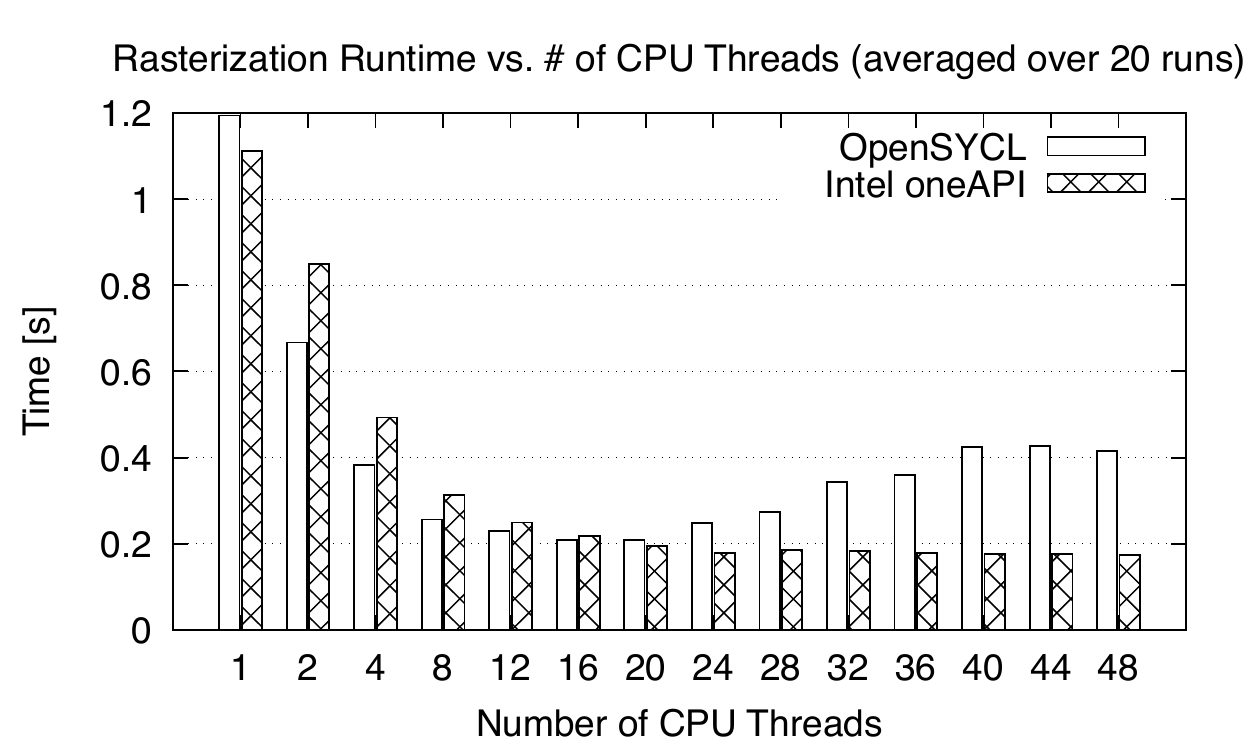} 
        \caption{Comparison of the Rasterization time on the CPU for the SYCL implementation with two different compilers.}\label{fig:sycl-cpu}
    \end{minipage}\hfill
    \begin{minipage}{0.48\textwidth}
        \centering
        \includegraphics[width=1.0\textwidth]{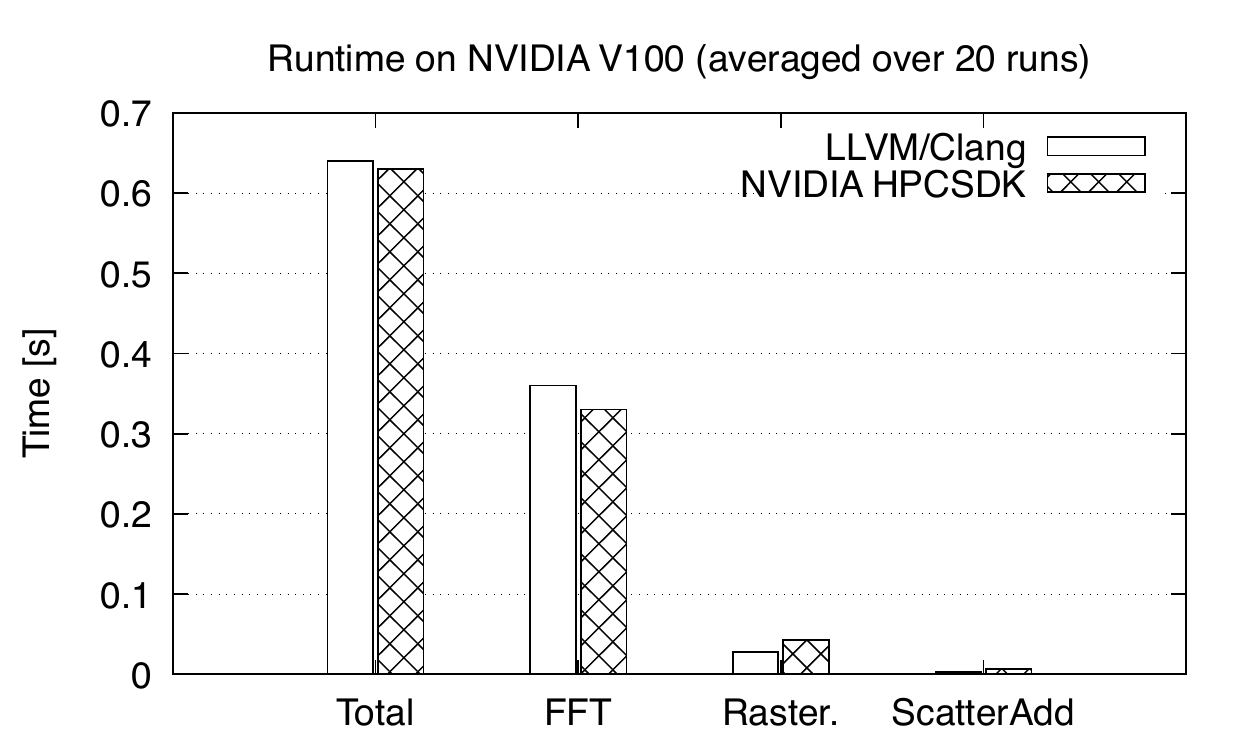} 
        \caption{Comparison of the LLVM/Clang compiler and the NVIDIA HPC SDK for the OpenMP implementation.}\label{fig:omp-nvidia}
    \end{minipage}
\end{figure}

Since we were able to compile the SYCL code for the CPU backend with two compilers, OpenSYCL and Intel oneAPI, we show the timing of the Rasterization kernel as a function of the number of CPU threads in Figure~\ref{fig:sycl-cpu}. OpenSYCL shows better scaling up to 16 threads, after which oneAPI outperforms OpenSYCL. This may be due to the different runtime libraries the compilers use, since OpenSYCL uses the OpenMP backend while oneAPI uses Intel TBB. 

For the OpenMP implementation, we were able to use both the LLVM/Clang compiler and the NVIDIA HPC SDK to compile the code for the NVIDIA V100 GPU. Both compilers give similar overall performance, as shown in Figure~\ref{fig:omp-nvidia}. 
\section{Summary and Outlook} \label{sec:summary}
We successfully implemented the LArTPC signal simulation in the Wire-Cell Toolkit with three portable programming models: Kokkos, SYCL and OpenMP. We have shown that all three programming models can produce comparable computational performance with the simulation parameters we tested. 
They all allow us to write a single code to target different computing architectures. We are in the process of investigating the \texttt{stdpar} model~\cite{stdpar} for WCT, and will report the detailed comparisons of these different programming models in a future publication. 

\ack 

This work was supported by the U.S. Department of Energy, Office of
Science, Office of High Energy Physics, High Energy Physics Center for
Computational Excellence (HEP-CCE) under B\&R KA2401045.  This research used resources of the National Energy Research Scientific Computing Center (NERSC), a U.S. Department of Energy Office of Science User Facility located at Lawrence Berkeley National Laboratory, operated under Contract No. DE-AC02-05CH11231. We gratefully acknowledge the support of the Wire-Cell team of the Electronic Detector Group in the Physics department and the Scientific Data and Computing Center of Brookhaven National Laboratory,  which is supported by the U.S. Department of Energy under Contract No. DE-SC0012704.

\section*{References}
\bibliography{refs}
\bibliographystyle{iopart-num}


\end{document}